\documentclass[conference]{IEEEtran}
\IEEEoverridecommandlockouts
\usepackage{cite}
\usepackage{amsmath,amssymb,amsfonts}
\usepackage{algorithmic}
\usepackage{graphicx}
\usepackage{textcomp}
\usepackage{comment}
\usepackage{float}
\usepackage{xcolor}
\def\BibTeX{{\rm B\kern-.05em{\sc i\kern-.025em b}\kern-.08em
    T\kern-.1667em\lower.7ex\hbox{E}\kern-.125emX}}
\begin{document}

\title{Statistical Characterization of Random Errors Present in Synchrophasor Measurements
\thanks{This work was supported in part by a Research Experiences for Undergraduates (REU) grant by the National Science Foundation under Award 1934766.}
}

\author{\IEEEauthorblockN{Demetra Salls, \emph{Undergraduate Member, IEEE}, Jairo Ramírez Torres, \emph{Undergraduate Member, IEEE}, Antos \\
Cheeramban Varghese, \emph{Graduate Member, IEEE}, John Patterson, \emph{Graduate Member, IEEE}, and Anamitra Pal, \\
\emph{Senior Member, IEEE}}
\IEEEauthorblockA{\text{School of ECEE} \\
\text{Arizona State University}
Tempe, AZ, USA}
}

\maketitle

\begin{abstract}
The statistical characterization of the measurement errors of a phasor measurement unit (PMU) is currently receiving considerable interest in the power systems community. This paper focuses on the characteristics of the errors in magnitude and angle measurements introduced only by the PMU device (called random errors in this paper), during ambient conditions, using a high-precision calibrator.
% to identify errors in the measurements of the PMU and the signals passed from the calibrator to the PMU. 
The experimental results indicate that the random errors follow a non-Gaussian distribution. They also show that the M-class and P-class PMUs have distinct error characteristics. The results of this analysis will help researchers design algorithms that account for the non-Gaussian nature of the errors in synchrophasor measurements, thereby improving the practical utility of the said-algorithms in addition to building on precedence for using high-precision calibrators to perform accurate error tests. 
\end{abstract}


\section{Introduction}
Phasor measurement units (PMUs) measure magnitude, phase angle, frequency, and rate-of-change-of-frequency (ROCOF) of electrical signals and play a critical role in real-time monitoring, protection, and control of modern power systems. 
% PMU measurements are used in various power system protection, monitoring and control applications. 
% These measurements, as expected with any real device, are not perfect. 
% The general assumption regarding the error distribution of the measurements obtained from these devices is that that they follow a Gaussian (normal) distribution. 
The measurements obtained from PMUs are affected not only by the PMU device mechanics, but also by other components in the measurement chain, such as instrument transformers and cables. As such, PMU errors can be of two types: \emph{systematic} and \emph{random} \cite{ahmad2019statistical}. Systematic errors occur due to the erratic behavior of the components present in the measurement chain (such as current transformer saturation), while random errors occur due to the PMU device itself (e.g., components internal to the PMU, response of the PMU to normal changes occurring in the system). The focus of this paper is on the \emph{random errors} of a PMU.

PMUs are being adopted rapidly by utilities and more power system applications are making use of them to achieve better performance \cite{9166750,10.1145/3397776.3397779}. 
% making the knowledge of the error distributions of these devices to become increasingly relevant to power engineers. 
% This paper is the product of the interest around understanding the ground truth of PMU measurement errors and the correlating fields of calibration and error testing. 
% The knowledge of the statistical characteristics of the PMU measurement errors will help researchers design better algorithms that account for their presence,
% % to handle noise and improve PMU error reporting,
% % thus providing better performance for various monitoring, protection, and control applications. 
% thereby improving the practical utility of the said-algorithms. 
Knowledge of the nature of the errors in PMU measurements is extremely important because by understanding their statistical characteristics,
% of the errors present in these measurements, 
researchers can design better algorithms that have more practical utility.  
The scope of this research is limited to characterizing the random errors that occur during \emph{ambient} conditions (i.e., in absence of faults, oscillations, time-synchronization issues, or device saturation), since a power system stays in this condition most of the time.

Traditionally, PMU measurement noise was always assumed to follow a Gaussian distribution. However, there has been some recent interest in understanding the exact statistical nature of PMU measurement noises. Brown et al. tried to quantify the PMU  measurement noises using field PMU data \cite{brown2016characterizing}. They conditioned the data, segregated the steady state data, and used filtering techniques to extract the errors. Studying the nature of these errors they concluded that it has a zero mean Gaussian distribution. Later, Wang et al., using redundant field PMU measurements and employing the fact that the difference of two random variables following a Gaussian distribution is a necessary condition for the individual random variables to be Gaussian, showed that PMU measurement errors do not necessarily follow a Gaussian distribution \cite{wang2017assessing}. Ahmad et al. further verified this argument using data driven filtering techniques \cite{ahmad2019statistical}. In particular, they used an adaptive moving window median absolute deviation method to extract the errors from both synthetic data as well as field data.
% used standard statistical methods to conclude
They concluded from their analysis
that the error followed a 3-component Gaussian mixture model (GMM) distribution, primarily due to device saturation. 

%However, both \cite{ahmad2019statistical} and \cite{wang2017assessing} concluded that the best way to characterize PMU errors is to use a high precision calibrator, and feed identical inputs into the calibrator and the PMU device-under-test (DUT) and compare the outputs.

In this paper, we have used a \emph{high precision Fluke calibrator} (Fluke 6135A) \cite{Fluke6135a_manual} to characterize the random errors found in PMU measurements. Although very expensive, a high precision calibrator is the best way to extract PMU measurement errors. This is because it precludes the assumption of presence of a calibrated PMU at every substation, which is employed in many state-of-the-art error extraction techniques  \cite{ahmad2019statistical, wang2017assessing}.
% They stated in their paper that a high precision calibrator is the best way to extract the errors for analysis \cite{ahmad2019statistical}. The authors of this paper have used a high precision calibrator (Fluke 6135A) to characterize random errors in PMU measurements.

% Due to the multitude of applications making use of PMU measurements and the significance of the accuracy of these applications in terms of monetary benefits, more detailed studies are required in this domain. This paper is one such study about the statistical characteristics of the random errors of PMU measurements.
The focus of this paper is on discovering error distributions during ambient system conditions. Therefore, the slowly changing dynamics of the power system as specified in the IEEE C37.118.1-2011 Standard \cite{2011_C37_118_std} were applied to the device-under-test (DUT). The analysis was carried out for a P-class PMU as well as an M-class PMU using normality tests as well as more quantitative metrics (higher-order moments). The results indicate that \emph{even under ambient conditions, the random errors in P-class and M-class PMUs have a distinct non-Gaussian error distribution}.

The rest of the paper is structured as follows. Section \ref{II} describes the mathematical basis for this work. Sections \ref{III} and \ref{IV} explain the test setup and results, respectively. The conclusion and future scope are identified in Section \ref{V}.

\section{Mathematical Background}
\label{II}
\subsection{Dynamic Tests}
In this research, three dynamic tests described in the IEEE C37.118.1-2011 Standard \cite{2011_C37_118_std}, namely, the amplitude modulation (AM) test, the phase modulation (PM) test, and the frequency ramp (FR) test, have been carried out, and the statistical analysis of the extracted errors have been reported. 

The signal for the AM and PM tests for the $a$-phase can be mathematically represented as:
  \begin{equation}
    X_a = X_m [1+k_x cos(\omega t)] cos[w_0 t + k_a cos(w t - \pi)]
\end{equation}
where, $k_x$ and $k_a$ represent the amplitude and phase modulation factors, respectively. For the $b$ and $c$ phases, the equation would be similar except for the fact that the phase would be lagging and leading by $\frac{2 \pi}{3}$, respectively.
The positive sequence signal can be obtained from the $abc$ phases.
% Combining these three the positive sequence signal can be found out.
If $T$ is  the phasor reporting interval and $n$ is any integer then the phasor representation of the positive sequence measurement obtained at $t=nT$ is given by:
\begin{equation}
\label{AM-PM-Phasor1}
    X(nT) = \{\frac{X_m}{\sqrt{2}}\} [1+k_x cos(\omega n T)] \angle\{k_a cos(\omega n T - \pi)\}
\end{equation}

The electrical signals generated in accordance with (\ref{AM-PM-Phasor1}), i.e. the \emph{true values}, are passed to the DUT. The values obtained at the output of the DUT are saved as the \emph{measured values}. 
% The error values can be calculated from the true and measured values.
The relative magnitude error (RME) is calculated as:
\begin{equation}
\label{AM-PM-RME}
\begin{aligned}
RME 
% \frac{X_{mag_{true}} - X_{mag_{meas}}}{X_{mag_{true}}}\\
%     RME &= \frac{(\{\frac{X_{m_{true}}}{\sqrt{2}}\}  - \{\frac{X_{m_{meas}}}{\sqrt{2}}\})  [1+k_x cos(\omega n T)]}{\{\frac{X_{m_{true}}}{\sqrt{2}}\} [1+k_x cos(\omega n T)]} \\
&= \frac{X_{m_{true}} - X_{m_{meas}}}{X_{m_{true}}}
\end{aligned}
\end{equation}

The phase angle error is given by:
\begin{equation}
\label{AM-PM-PE}
\begin{aligned}
PE
&= X_{ang_{true}} - X_{ang_{meas}}\\
%   &= (\omega_{true}nT +  k_a cos(\omega_{mod} n T - \pi)) \\
%   &- (\omega_{meas}nT + k_a cos(\omega_{mod} n T - \pi)\\
% &=  (\omega_{true} - \omega_{meas})nT
\end{aligned}
\end{equation}

\begin{comment}
The corresponding measured signals is given by:
\begin{equation}
    X_{meas}(nT) = \{\frac{X_{m_{meas}}}{\sqrt{2}}\} [1+k_x cos(\omega n T)] \angle\{k_a cos(\omega n T - \pi)\}
\end{equation}
\end{comment}

% The third type of dynamic tests that have been carried out is the frequency ramp tests. 
For the FR tests, the positive sequence signal is written as:
\begin{equation}
    X_1 = X_m cos(w_0 t + \pi R_f t^2)
\end{equation}
where, $R_f= \frac{df}{dt}$. At reporting time tags, $t=nT$, the phasor measurements can be expressed as:
\begin{equation}
    X(nT) = \frac{X_m}{\sqrt{2}}\angle\{ \pi R_f (nT)^2\}
\end{equation}

%%%% Lines 133-145 are commented out because the frequency errors have not been reported in the paper

\begin{comment}
The frequency errors  
% and ROCOF error in these cases 
can now be found as:
\begin{equation}
    \Delta f(nT) = R_f nT
\end{equation}
% \begin{equation}
%   \frac{d}{dt}[f(nT)] = R_f
% \end{equation}

The relative magnitude error, phase angle error and frequency error corresponding to these tests are calculated for both P-class and M-class PMUs. 
\end{comment}

The relative magnitude error and phase angle error corresponding to these tests are calculated for both P-class and M-class PMUs. Note that the relative magnitude error is the ratio of the magnitude error and the true value (reported as a percentage value), while the phase angle error is the difference between the measured value and the true value (reported in degrees). Finally, the histograms of these errors are plotted to draw inferences regarding the distribution of the random errors, while numerical and statistical tests are performed to quantify the parameters of the distribution. 
% if there is a Gaussian distribution, or for that matter any other distribution, that can accurately explain the error distributions.

 \subsection{Normality Tests}
Normality tests are primarily used to verify if a data set can be reasonably modeled by a Gaussian distribution or not. Two normality tests that are used in this study are the Shapiro-Wilks test and the Kolmogorov-Smirnov test.
 
 The Shapiro-Wilks test operates by testing the data set across a hypothesis of normality, creating a $p$-value that, within a certain $alpha$ significance level, indicates whether the researchers can accept or reject the null hypothesis that the data fits the normal distribution.  
The critical value of $alpha$ is a cut-off point where the experiment may reject a null hypothesis in error. Common values of $alpha$ used in practice are 0.1, 0.05, and 0.01, respectively \cite{Engineering_Statistics_Handbook_Alpha}. For this experiment, the value of $alpha$ was set at 0.05, indicating that the probability of a \emph{Type 1} error (falsely rejecting a null hypothesis that is actually true) is less than 5\%.
 
 The Kolmogorov-Smirnov test 
%  is the other type of normality test used and it
 is based on the empirical cumulative distribution function (ECDF). The null hypothesis in this case is that both sets of input data are coming from the same distribution \cite{dallal1999little}. Thus, the Kolmogorov-Smirnov test calculates the distance between ECDFs of the given error distribution and corresponding Gaussian distribution to determine the similarity between the two. The next section describes the test setup created for this study.

\section{Hardware Test Setup}
\label{III}
In order to provide accurate time-synchronized instrument transformer-level signals to the DUT,
% for use in the characterization of their error distributions, 
a Fluke 6135A PMU calibration unit was employed. This hardware device generates time-synchronized electrical waveforms of voltage and current that are guaranteed
% by the manufacturer 
to meet 0.007\% accuracy for the tests specified in \cite{Fluke6135a_manual}. Each DUT was connected to the Fluke 6135A such that its voltage and current input terminals were wired to the voltage and current output terminals of the Fluke 6135A. Fig. 1 depicts one of the DUTs (an M-class PMU) connected to the Fluke 6135A.

\begin{figure}
\centering
\includegraphics[width=0.48\textwidth]{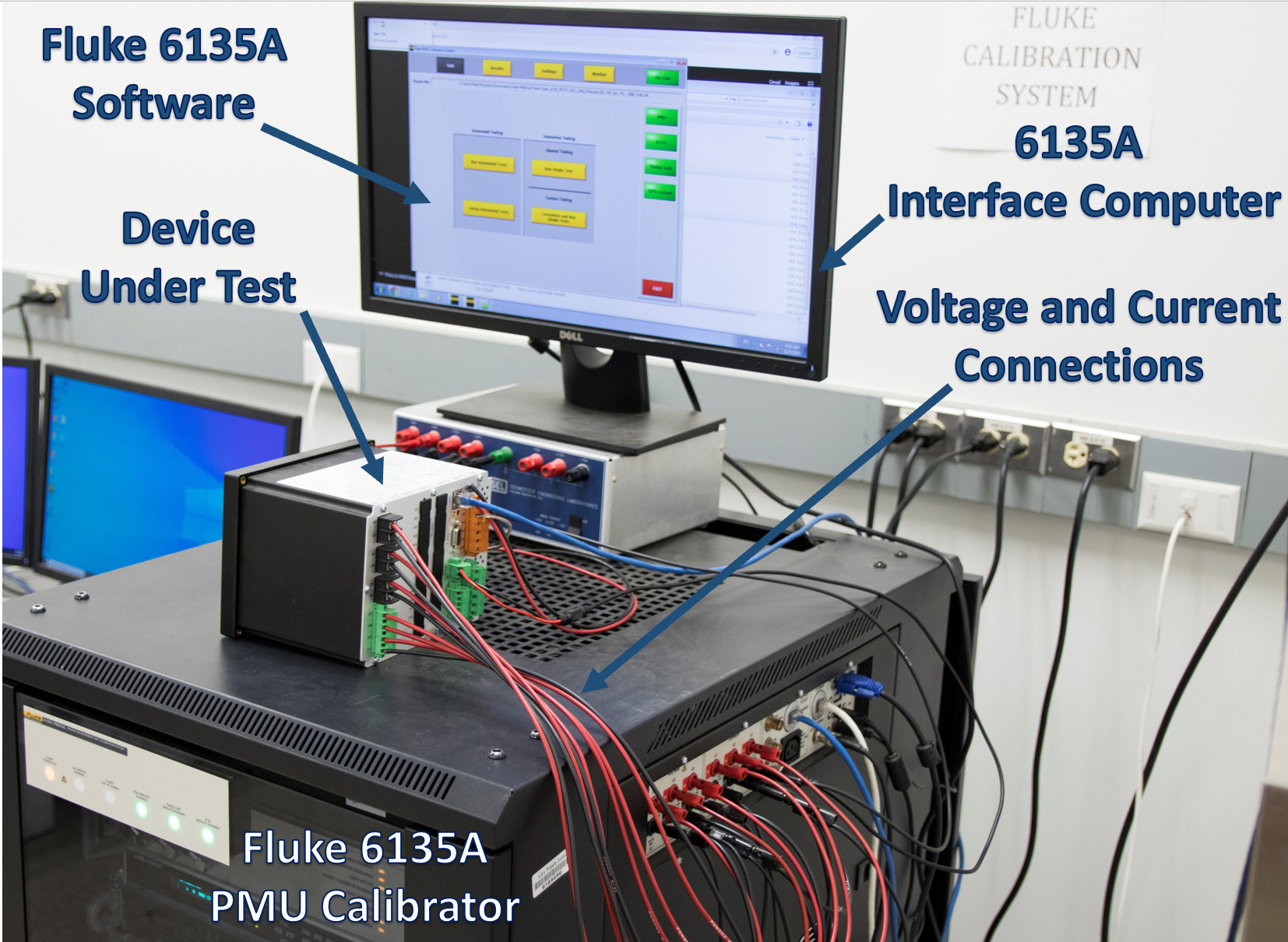}
\caption{Hardware Test Setup}
\label{fig:medium_line}
\end{figure}

The Fluke 6135A PMU calibrator can be configured to perform all the steady-state and dynamic performance tests required by the IEEE C37.118-2011 \cite{2011_C37_118_std}. When each test is performed by the calibrator, a spreadsheet of synchrophasor measurements and their respective nominal values (as produced by the calibrator at its output terminals) are reported by the software suite provided with the calibrator. These test report spreadsheets also include the errors in angle and magnitude between the measured and the nominal synchrophasor, and it is these error values that were extracted from each test to characterize the errors in the DUT.

The goal of this characterization experiment is to determine the behavior of the PMUs under dynamic conditions that are reflective of the behavior of an actual power system during ambient conditions. The IEEE C37.118.1a-2014 Standard \cite{IEEE_C37_118_2014a} specifies the parameters for the dynamic tests for sinusoidal AM, sinusoidal PM, and linear frequency ramping. 
% In the case of the amplitude and phase modulation tests, a minimum, middle, and maximum modulation frequency test was selected to be performed. 
In the case of the AM and PM tests, the minimum modulation frequency was employed.
% as it was most representative of the behavior of a normal power system. 
For the FR tests, a gradually rising frequency ramp test and a gradually falling frequency ramp test were selected. In each case, the test parameters were selected as specified in the IEEE C37.118.1a-2014 Standard \cite{IEEE_C37_118_2014a}; the numerical values of the parameters are given in Section \ref{IV.B}.
In order to extract sufficient data to plot histograms of the error distributions of interest for each test, the relevant tests were performed repeatedly until a total of 18,000 samples (10 minutes worth of test data at 30 samples per second) were collected. These error values were concatenated together, plotted, and checked for consistency. The latter was ensured by repeating every test three times. Afterwards, the most relevant trends and characteristics of the errors were noted and conclusions were drawn on their basis.

% In order to extract sufficient data to plot detailed histograms of the error distributions of interest for each test, the relevant tests were each performed repeatedly until a total of 18,000 samples (10 minutes worth of test data) were collected. These error values were concatenated together, plotted, and checked for consistency. Once consistency was verified, the most relevant trends and characteristics of the errors were noted and conclusions were drawn from them.

%There are three types of dynamic tests from the list of tests specified in the IEEE C37.118.1-2011 standard. For each of amplitude modulation, phase modulation and frequency ramp tests were performed for two extreme cases and an intermediate cases for both M-class and P-class PMUs. To be precise the amplitude and phase modulations were conducted for frequency of variation (of amplitude/ phase) as 0.1, 2.5 and 5.0 Hz for the M-class device. The same tests were carried out for frequencies 0.1, 0.9 and 1.9 Hz for the P-class device.

\section{Results}
\label{IV}
\subsection{Observations}
Some general observations made from the experiments are summarized below:
\begin{itemize}
    \item For the P-class PMU, when running the PM tests, the phase angle errors were found to be continuous, while the relative magnitude errors were found to be discrete (here, discrete means that the error values were placed into distinct bins rather than being continuously spread).
    \item For the P-class PMU, when running the AM tests, the relative magnitude errors were found to be continuous, while the phase angle errors were discrete.
    %\item As stated earlier, Gaussian (normal) distributions are only appropriate when data is a continuous variable.
    \item For both the P-class as well as the M-class PMU, the error distributions appeared to have a constant bias (a non zero mean value) for all the tests.
    \item The Shapiro-Wilks and the Kolmogorov-Smirnov tests substantiated the hypothesis that the random errors in both P-class and M-class PMU measurements have non-Gaussian distributions.
    %%\item The mean, standard deviation, skewness, and kurtosis values of each distribution are calculated to confirm both visual indicators and results of the Shapiro-Wilks and Kolmogorov-Smirnov tests accuracy with numerical evidence. 
   %% \item High values of skewness indicate that the error distribution is not symmetric, either right-tailed or left-tailed.
   %% \item Kurtosis values that have a large magnitude (absolute value), indicate that the distribution is deviant from that of a normal distribution in the sense that the tails are either less extreme (negative) or more extreme (positive) than those of a normal Gaussian distribution.
    %\item .
\end{itemize}

\subsection{Error Characteristics for the P-class PMU}
\label{IV.B}

\begin{figure}[H]
\centering
\includegraphics[width=.35\textwidth]{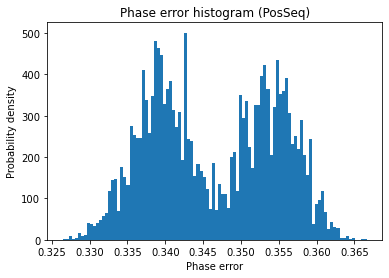}
\caption{P-class - phase angle error histogram - phase modulation}
\label{GE_PM_PSP}
\end{figure}

The error characterization studies were conducted for each of the three phases as well as the positive sequence.
% While the analysis was done for each of these cases , for
For sake of brevity, only the results of positive sequence tests corresponding to 
% the most realistic 
ambient power system operating conditions (slowly changing system dynamics) are discussed here. 
% The results for the other cases reflected similar inferences.
The following histograms correspond to the 0.1 Hz PM, 0.1 Hz AM, up to +0.03 Hz/sec FR up, and up to -0.03 Hz/sec FR down, tests \cite{2011_C37_118_std}. 
% The positive sequence error histograms of phase error and relative magnitude error are shown below. 
Fig. \ref{GE_PM_PSP} shows the phase angle error distribution for the PM tests. By repeating the tests three times it was confirmed that the shapes and ranges of the histograms were consistent. Two distinct peaks are visible in this error histogram, indicating that it can be more accurately represented using a two-component GMM as opposed to a Gaussian distribution.

\vspace{-3mm}

\vspace{-2mm}

\begin{figure}[H]
\centering
\includegraphics[width=.35\textwidth]{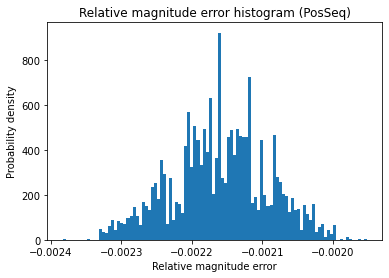}
\caption{P-class - relative magnitude error histogram - amplitude modulation}
\label{GE_AM_PSM}
\end{figure}

Fig. \ref{GE_AM_PSM} shows the relative magnitude error histogram of the same device for AM tests. The normality tests and the skewness \& kurtosis parameters (calculated during the quantitative analysis in Section \ref{IV.D}) reinforce the hypothesis that the relative magnitude error distributions are non-Gaussian for this case as well.

\vspace{-5mm}

\vspace{-6mm}

\begin{figure}[H]
\centering
\includegraphics[width=.35\textwidth]{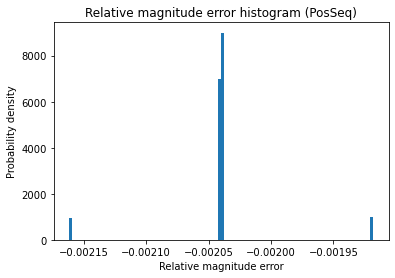}
\caption{P-class -  relative magnitude error histogram -  frequency ramp up}
\label{GE_FR_PSM}
\end{figure}

\begin{figure}[H]
\centering
\includegraphics[width=.35\textwidth]{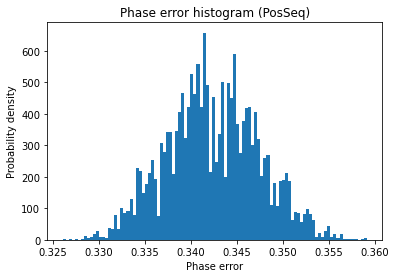}
\caption{P-class - phase angle error histogram - frequency ramp up}
\label{GE_FR_PSP}
\end{figure}

Figs. \ref{GE_FR_PSM} and \ref{GE_FR_PSP} represent the relative magnitude error histogram and phase angle error histogram for the FR up test. The relative magnitude error histogram shows discrete values with change in frequencies.
%The phase angle error histogram shows a distribution which has a heavier tail towards the right. 
The phase angle error histogram 
% for the frequency ramp on the other hand 
has error values distributed continuously. 
% in every bins present in the range.
The normality tests and the skewness and kurtosis values (see Section \ref{IV.D}) further validate the non-Gaussian nature of this error distribution.

\vspace{-6mm}

\subsection{Error Characteristics for the M-Class PMU}
\begin{figure}[H]
\centering
\includegraphics[width=.35\textwidth]{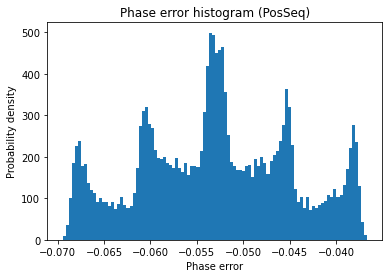}
\caption{M-class - phase angle error histogram - phase modulation}
\label{SEL_PM_PSP}
\end{figure}

The same set of tests (although with slightly different sets of parameters as mentioned in the IEEE C37.118.1-2011 Standard \cite{2011_C37_118_std}) when applied to the M-class PMU also resulted in non-Gaussian histograms for the random errors. Fig. \ref{SEL_PM_PSP} represents the phase angle error histogram for the PM test. From the histogram, five distinct peaks can be observed (as opposed to two in the case of the P-class PMU). As expected, the normality tests confirmed that the distribution was non-Gaussian. It was found that this error distribution could be best approximated by using a five-component GMM or a five-component beta mixture model. 
% A three-component GMM would also be a better approximation for this error characteristic than a simple Gaussian distribution. 

\begin{figure}[H]
\centering
\includegraphics[width=.35\textwidth]{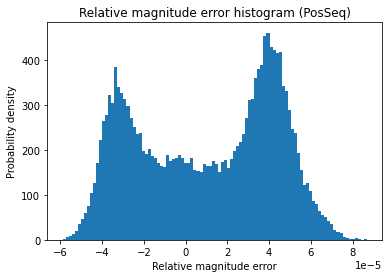}
\caption{M-class - relative magnitude error histogram - amplitude modulation}
\label{SEL_AM_PSM}
\end{figure}

The relative magnitude error histogram for AM test for the same device is shown in Fig. \ref{SEL_AM_PSM}. Even though visual inspection of this histogram reveals that it is indeed a non-Gaussian distribution with two-peaks, since the range is extremely small, this distribution may not have much significance in practice.

\begin{figure}[H]
\centering
\includegraphics[width=.35\textwidth]{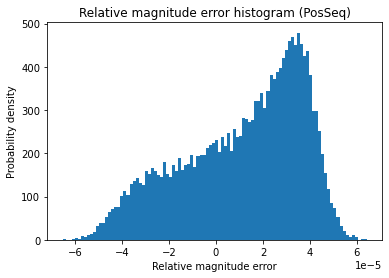}
\caption{M-class - relative magnitude error histogram - frequency ramp down}
\label{SEL_FR_PSM}
\end{figure}

\begin{figure}[H]
\centering
\includegraphics[width=.35\textwidth]{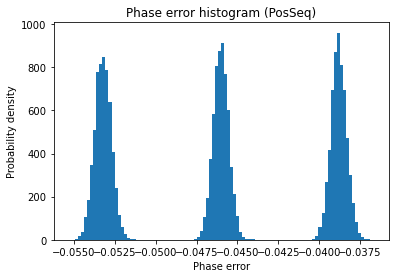}
\caption{M-class - phase angle error histogram - frequency ramp down}
\label{SEL_FR_PSP2}
\end{figure}

The relative magnitude error histogram for FR down test is shown in Fig. \ref{SEL_FR_PSM}. It can be observed that this histogram is asymmetric with a heavier tail towards the left side. The phase angle error histogram for the FR down test is shown in Fig. \ref{SEL_FR_PSP2}. The error histogram can be best modeled using a tri-modal distribution as the individual components are distinct.

\vspace{-6mm}

\begin{comment}
Table I compares the random error characteristics of P-class and M-class PMUs. It appears that the P-class and M-class PMUs have significantly different error characteristics and they need to be studied separately.

\begin{table}[H]
\caption{Comparison of error characteristics of M-class and P-class PMUs}
\begin{tabular}{|l|l|l|}
\hline
 & P-class & M-class  \\ \hline
 PSP - PM& \includegraphics[width=.15\textwidth]{GE_PM_IPS_PhaseErr.png}  & \includegraphics[width=.15\textwidth]{SEL_PM_IPS_PhaseErr.png} \\ \hline
PSP - AM& \includegraphics[width=.15\textwidth]{GE_AM_IPS_RelErr.png}  &  \includegraphics[width=.15\textwidth]{SEL_AM_IPS_RelErr.png} \\ \hline
PSP - FR-U& \includegraphics[width=.15\textwidth]{GE_Up_FR_IPS_PhaseErr.png}  & \includegraphics[width=.15\textwidth]{SEL_Up_FR_IPS_PhaseErr.png}  \\ \hline
PSP - FR-D & \includegraphics[width=.15\textwidth]{GE_Down_FR_IPS_PhaseErr.png} & \includegraphics[width=.15\textwidth]{SEL_Down_FR_IPS_PhaseErr.png} \\ \hline
\end{tabular}
\end{table}

\end{comment}

\subsection{Quantitative analysis}
\label{IV.D}
In addition to the normality tests, some statistical quantities which determine the features of a probability distribution were also calculated and used to verify the non-Gaussian nature of the random errors in PMU measurements. The mean, standard deviation, skewness, and kurtosis are the first, second, third, and fourth moments, respectively, and convey specific attributes of a distribution. The mean and standard deviation represent the average value and the range of the distribution. The skewness of the distribution indicates if the error distribution is asymmetric (either right-tailed or left-tailed); i.e., for a Gaussian distribution the skewness value is negligibly small (ideally zero). The kurtosis denotes the amount of data present towards both ends relative to the center of the distribution; this should also be negligible (ideally zero) for a Gaussian distribution. 
% The four moments calculated for PMU measurement errors thus can convey different characteristics. 

Tables \ref{Table I} and \ref{Table II} show the phase angle errors and the relative magnitude errors for the P-class PMU, respectively. PM, AM, and FR represent the three dynamic tests, namely, phase modulation, amplitude modulation and frequency ramp.
The relatively large values of skewness and kurtosis in these tables are proof of the non-Gaussian nature of the random errors in PMU measurements during ambient conditions.

\begin{table}[H]
\centering
\caption{Phase Angle Error - P-class PMU }
\label{Table I}
\begin{tabular}{|l|l|l|l|l|l|}
\hline
   & Mean     & Median   & Std. dev.      & Skewness & Kurtosis   \\ \hline
PM & 0.361392 & 0.36019  & 0.00779 & 0.077721 & -1.426     \\ \hline
AM & 0.344971 & 0.34570 & 0.00918 & -0.01066 & -1.26174  \\ \hline
FR & 0.342382 & 0.342188 & 0.00504 & 0.068226 & -0.31194 \\ \hline
\end{tabular}
\end{table}

\vspace{-6mm}

\begin{table}[H]
\centering
\caption{Relative Magnitude Error - P-class PMU }
\label{Table II}
\begin{tabular}{|l|l|l|l|l|l|}
\hline
   & Mean     & Median   & Std. dev.      & Skewness & Kurtosis   \\ \hline
PM & -0.00211 & -0.00216 & 5.97E-05 & 0.43024  & -1.79      \\ \hline
AM & 0.342005 & 0.340205 & 0.00443 & 0.18960 & -0.22678 \\ \hline
FR &-0.00201 & -0.00204 & 5.45E-05 & 1.00192 & -0.8246  \\ \hline
\end{tabular}
\end{table}

Tables \ref{Table III} and \ref{Table IV} denote the phase angle errors and the relative magnitude errors for the M-class PMU, respectively.
These tables validate the following observations: (a) the non-Gaussian nature of the error distributions, and (b) the existence of bias in the error values.

\begin{table}[H]
\centering
\caption{Phase Angle Error - M-class PMU }
\label{Table III}
\begin{tabular}{|l|l|l|l|l|l|}
\hline
   & Mean   & Median & Std. dev.     & Skewness & Kurtosis \\ \hline
PM & -0.053 & -0.053 & 0.00818 & 0.00592  & -0.747   \\ \hline
AM & -0.046 & -0.046 & 0.00591 & 0.00724  & -1.48    \\ \hline
FR &  -0.04598 & -0.045986 & 0.00589 & -0.00910    &  -1.47634   \\ \hline
\end{tabular}
\end{table}

\vspace{-5mm}

\begin{table}[H]
\centering
\caption{Relative Magnitude Error - M-class PMU }
\label{Table IV}
\begin{tabular}{|l|l|l|l|l|l|}
\hline
   & Mean     & Median   & Std. dev.      & Skewness & Kurtosis \\ \hline
PM & 1.55E-05 & 1.55E-05 & 7.51E-06 & 0.0178   & -0.0101  \\ \hline
AM & 1.29E-05 & 1.78E-05 & 3.32E-05 & -0.152   & -1.33    \\ \hline
FR & 1.15E-05 & 1.73E-05  & 2.56E-05  & -0.56436    & -0.68725    \\ \hline
\end{tabular}
\end{table}

\subsection{Discussion}
% From the dynamic tests it was observed that the phase angle error value consistently varied more than the relative magnitude error values. 
For the PM test on the M-Class PMU, the relative magnitude errors were not only minuscule, but also across all relative magnitude errors the error distributions did not have a large variance (the standard deviation ranged from the order of $10^{-6}$ to $10^{-4}$).   This observation makes sense intuitively because M-Class PMUs are expected to give highly precise measurements of those quantities that are not varied (the amplitude is held constant during the PM test).
% , indicating the high accuracy and general Gaussian shape).
% For the other cases, the relative magnitude error values are also significant and all of them show non-Gaussian behaviors.  
% Phase angle error values are larger numerically, therefore, the effect of random errors are more pronounced in the case of phase angle errors. 
% It is also interesting to note that while the numerical values of relative magnitude errors are smaller, the variations and the moments of their distributions have higher values. Conversely, while the numerical values of the phase angle errors are relatively large, the variations (the second, third, and fourth moments) are small. 

% In power system operations even a phase angle measurement error of $10^{-1}$ can have significant impact on the resulting performance of the algorithms that use it. Therefore, it is important to consider the characteristics of PMU measurement errors with as much accuracy as possible. Having the knowledge of the parameters of their probability distributions definitely helps in this regard.  
The P-class PMU was found to be less accurate than the M-class PMU. This is also expected because of the different functions that the two devices serve; the P-class PMU is primarily designed for protection, and hence values speed over accuracy, while the M-class PMU is primarily designed for high quality measurements, and hence values accuracy over speed.   
Having said that, the errors in both PMU classes can be significantly different even under similar system conditions. Therefore, the impact that these errors can have on the algorithms that use measurements from these two types of PMUs must be carefully considered.  

% and impact the power measurements and should be considered when functioning in future repetitions. 
%Reiterating the key purpose of this experiment, however, is that the non-Gaussian nature of these random error distributions is evident when looking numerically at both the skewness and, especially, the kurtosis values as well as the visual depictions of the error distributions. Cases across both the M-class and P-class devices have kurtosis values with a magnitude larger than 1, indicating the curves are not ideal Gaussian-errors.  The histograms indicate that possibly a 5-component GMM or beta mixture model would be more apropos for some curves while others would suggest a 2-component GMM distribution, both of which would support the rejection of the hypothesis that the distribution of errors present in a Gaussian manner. 

\section{Conclusion}
\label{V}
The statistical characterization of the random errors present in the magnitude and angle measurements of P-Class and M-Class PMUs showed that their distribution is not necessarily Gaussian even during ambient system conditions. 
%There are some cases as stated earlier where a Gaussian curve does indeed prove to be the best fitting distribution; however, it is not accurate for all modulation tests. 
This was the case when any of the quantities, namely, phase, amplitude, or frequency, was varied individually. A non-zero bias was also observed in the random errors.  
The most realistic scenario of power system operation is one where amplitude, phase, and frequency vary simultaneously (within  specific bounds). This scenario will be explored in the future. 
% Understanding the nature of Frequency errors and ROCOF errors are crucial in PMU error characterization studies. This will also be studied in detail.
The errors in frequency and ROCOF will also be explored in a future study.
% along with the determination of whether the other distributions can be properly categorized. 
Another topic for future research is the study of the systematic errors in PMU measurements. The combined characterization of the random errors and the systematic errors using a high-precision calibrator will provide an even more realistic understanding of the characteristics of PMU measurement errors.

%valuable insights into the ground truths of PMU measurement errors in the field.  Because this research looked solely at random errors due to the functions within a PMU, that data can be extrapolated to numerous cases; learning systematic errors, meanwhile, allow for a greater understanding of errors within specific environments and testing situations, opening up further study into more particular scenarios.  

%This study has been able to not only provide strong evidence that would suggest PMU error measurements do not present in a Gaussian distribution, but also open further avenues into research available to ensure that PMUs remain a stable measurement tool within the power industry.  

\vspace{1pt}

\bibliographystyle{ieeetr}
\bibliography{conference_041818}

\end{document}